\documentclass[aps,prl,twocolumn,a4paper,superscriptaddress]{revtex4}

\pdfoutput=1

\usepackage{multibbl}
\usepackage[pdftex]{graphicx}
\usepackage{amsmath,SIunits}
\usepackage[latin1]{inputenc}
\usepackage{latexsym,stmaryrd}
\usepackage[sort&compress]{natbib}

\bibpunct{}{}{,}{s}{}{,}

\makeatletter
\newcommand*{\citenst}[2][]{%
  \begingroup
  \let\NAT@mbox=\mbox
  \let\@cite\NAT@citenum
  \let\NAT@space\NAT@spacechar
  \let\NAT@super@kern\relax
  \renewcommand\NAT@open{[}%
  \renewcommand\NAT@close{]}%
  \citep{#2}%
  \endgroup
}
\makeatother

\begin{document}

\title{An Anderson-localized random nanolaser}

 \author{J.~Liu}
 \affiliation{DTU Fotonik, Department of Photonics Engineering, Technical University of Denmark, {\O}rsteds Plads 343, DK-2800 Kgs.~Lyngby, Denmark}
 \affiliation{Niels Bohr Institute, University of Copenhagen, Blegdamsvej 17, DK-2100 Copenhagen, Denmark}
 \author{P.~D.~Garcia}
 \affiliation{Niels Bohr Institute, University of Copenhagen, Blegdamsvej 17, DK-2100 Copenhagen, Denmark}
 \author{S.~Ek}
 \author{N.~Gregersen}
 \author{T.~Suhr}
 \author{M.~Schubert}
 \thanks{Present address: Universität Konstanz, Fachbereich Physik, Fach M 700, 78457 Konstanz, Germany.}
 \author{J.~M{\o}rk}
 \affiliation{DTU Fotonik, Department of Photonics Engineering, Technical University of Denmark, {\O}rsteds Plads 343, DK-2800 Kgs.~Lyngby, Denmark}
 \author{S.~Stobbe}
 \author{P.~Lodahl}
 \email{lodahl@nbi.ku.dk}
 \homepage{www.quantum-photonics.dk}
 \affiliation{Niels Bohr Institute, University of Copenhagen, Blegdamsvej 17, DK-2100 Copenhagen, Denmark}

\date{\today}

\small

\begin{abstract}
Precision is a virtue throughout science in general and in optics in particular where carefully fabricated nanometer-scale devices hold great promise for both classical and quantum photonics.~\cite{Baba,Vucovick,Lukin,Jahnke,Johnson,Painter} In such nanostructures, unavoidable imperfections often impose severe performance limits but, in certain cases, disorder may enable new functionalities.~\cite{Sapienza2010} Here we demonstrate on-chip random nanolasers where the cavity feedback is provided by the intrinsic disorder in a semiconductor photonic-crystal waveguide, leading to Anderson localization of light.~\cite{Anderson} This enables highly efficient and broadband tunable lasers with very small mode volumes. We observe an intriguing interplay between gain, dispersion-controlled slow light, and disorder, which determines the cross-over from ballistic transport to Anderson localization. Such a behavior is a unique feature of non-conservative random media that enables the demonstration of all-optical control of random lasing. Our statistical analysis shows a way towards ultimate thresholdless random nanolasers.
\end{abstract}

\maketitle

\small

\begin{figure}[t!]
\centering
\includegraphics[width=\columnwidth]{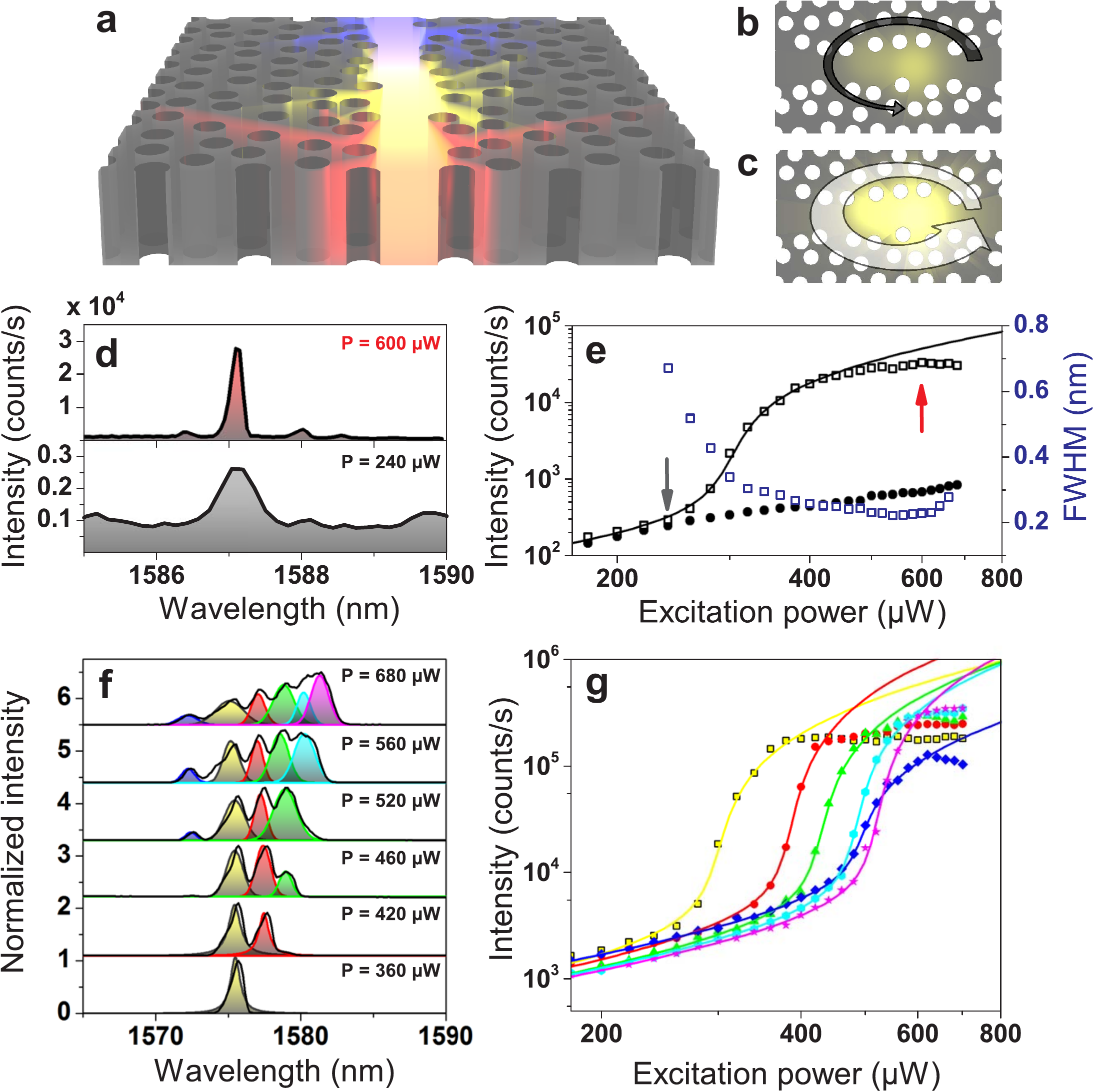}
\caption{\textbf{Single-mode and multi-mode random lasing in the Anderson-localization regime.} \textbf{a}, Illustration of a random laser in a disordered photonic-crystal waveguide formed from Anderson-localized modes, where the different colors indicate that multifrequency lasing can occur. \textbf{b}, and \textbf{c}, Illustration of multiple-scattering process below and above threshold, respectively. Below threshold, loss is dominating and Anderson-localized cavities do not build up. Above threshold, gain compensates loss and Anderson-localized random lasing sets in. \textbf{d}, Measured photoluminescence spectra of a single Anderson-localized lasing mode below (lower plot) and above (upper plot) the lasing threshold. \textbf{e}, Output intensity (black squares) and linewidth (blue squares) measured vs.~excitation power. The solid line represents a fit to semiconductor-laser rate equations and the black dots represent the background quantum-well fluorescence displaying no lasing. The arrows indicate the excitation power of the spectra shown in \textbf{d}. \textbf{f}, Photoluminescence spectra vs.~excitation power in multi-mode operation where the different colored curves are fits with Lorentzian functions representing different lasing modes. \textbf{g}, Output intensity vs.~excitation power corresponding to the different lasing modes shown in \textbf{f}. The solid lines represent the fit to semiconductor-laser rate equations for each lasing mode.}
\label{random_lasing}
\end{figure}

\begin{figure}
\centering
\includegraphics[width=8cm]{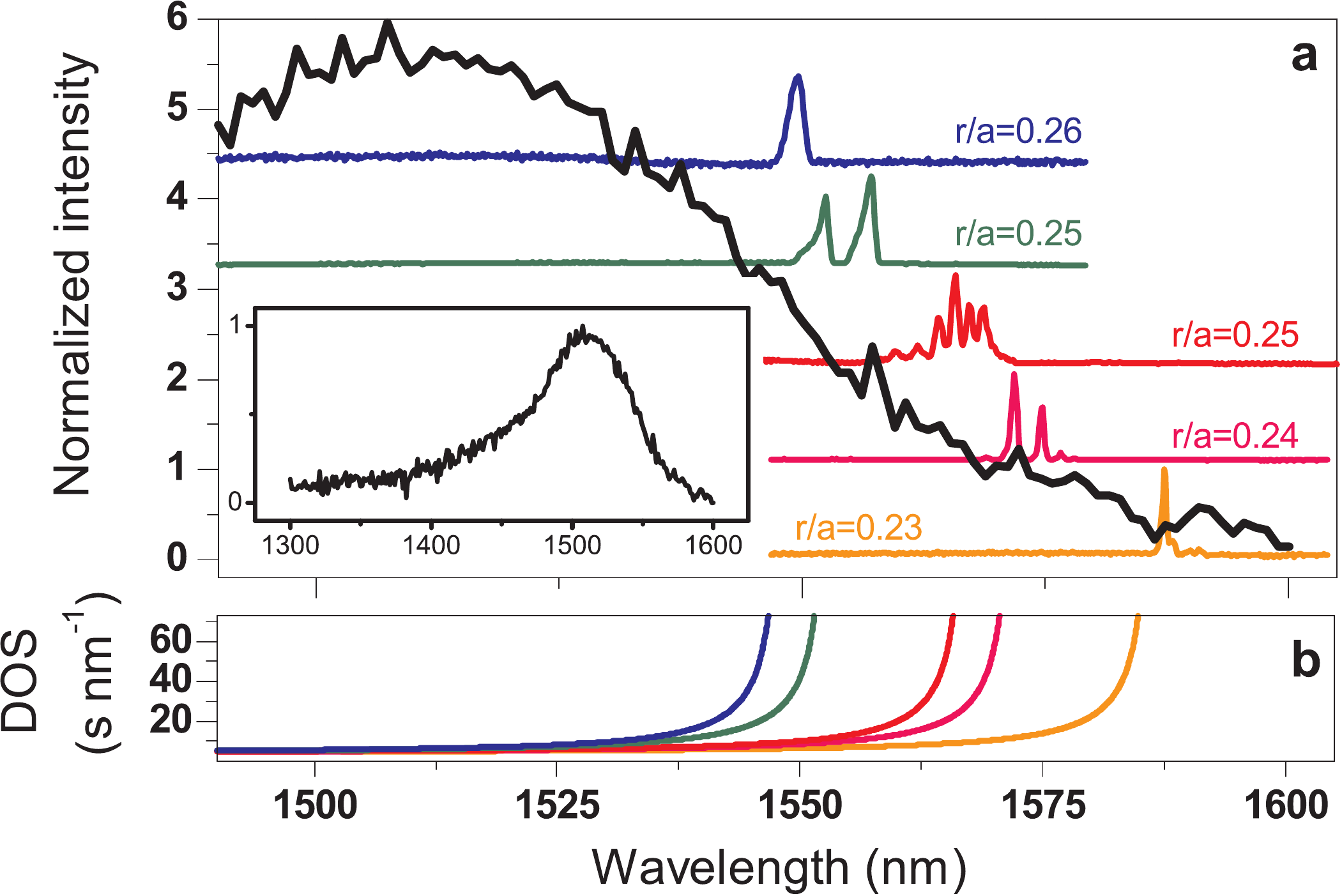}
\caption{\textbf{Tunable Anderson-localized random lasing.} \textbf{a}, Photoluminescence spectra (vertically shifted for visual clarity) above lasing threshold collected for various values of $\textit{r}/\textit{a}$, where $r$ is the hole radius and $a$ the lattice constant of the photonic crystal. The quantum well emission spectrum is shown for comparison (black line and inset). \textbf{b}, Calculated density of states of a photonic-crystal waveguide without disorder where the different colors correspond to the different values of $\textit{r}/\textit{a}$ shown in \textbf{a}.}
\label{Tunability}
\end{figure}

\begin{figure*}
\centering
\includegraphics[width=17cm]{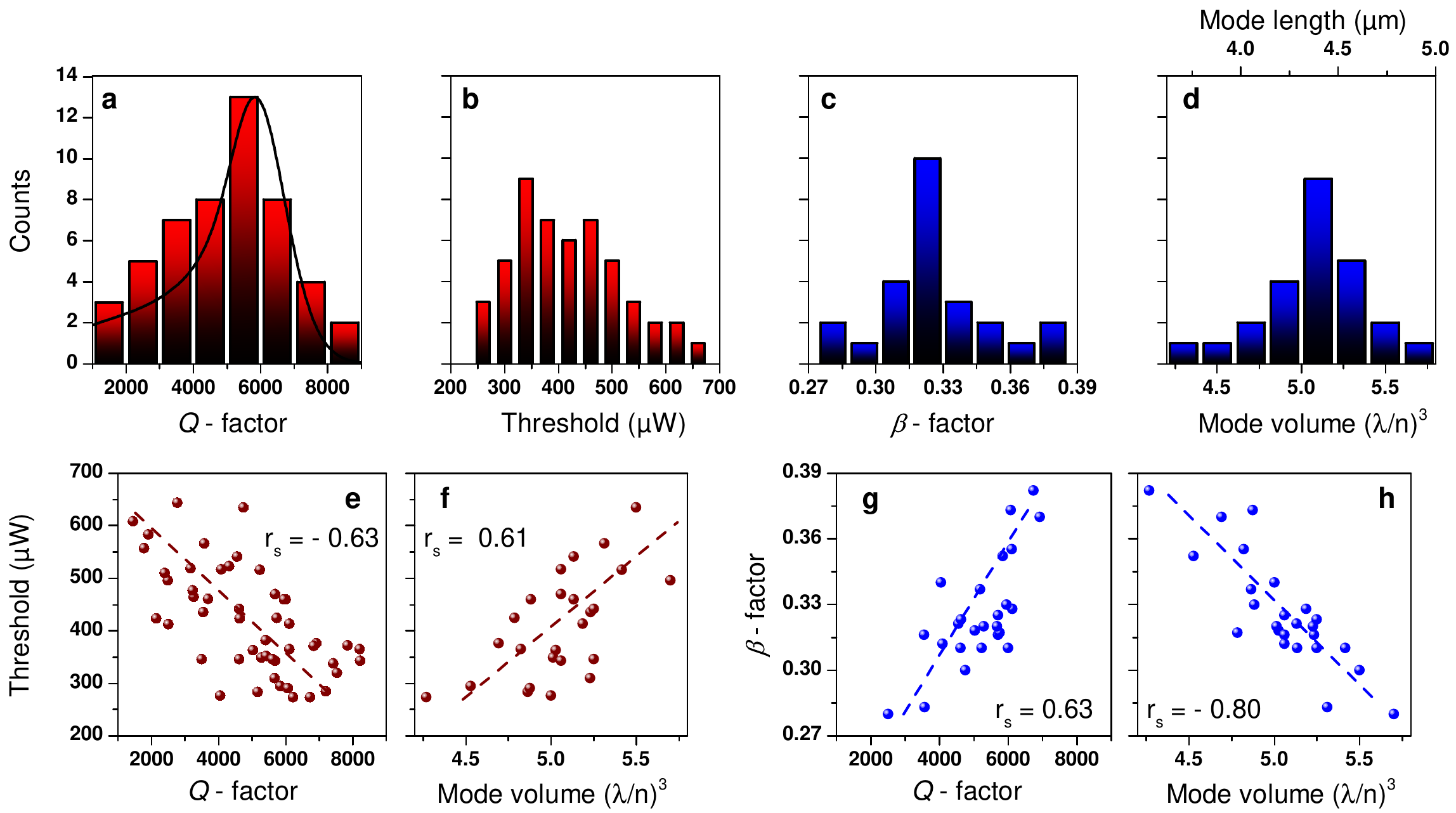}
\caption{\textbf{Statistical properties of random lasing in the Anderson-localization regime.} \textbf{a}, Experimental $Q$-factor distribution of the observed localized modes in the photonic-crystal waveguide at the threshold power (histograms) fitted with a 1D multiple-scattering model to extract the localization length (black line). \textbf{b}, Laser-threshold distribution obtained from measured input-output curves. \textbf{c}, $\beta$-factor and \textbf{d}, mode-volume (converted to mode length on the top axis) distributions extracted from measured input-output curves. \textbf{e},\textbf{f} and \textbf{g},\textbf{h} Measured threshold and $\beta$-factor versus  $Q$-factor and mode volume, respectively. The Spearman correlation coefficient, $\text{r}_{\mathrm{s}}$, describes the correlation of the different parameters, with 1 corresponding to perfect correlation and -1 to perfect anti-correlation. The dashed lines are guides to the eye.}
\label{Statistics}
\end{figure*}

The last decade has witnessed important breakthroughs in the search for the ultimate nanolaser with zero threshold and ultra-small size~\cite{Jahnke,ultimate_nanolaser}. This quest involves perfecting the nanocavity that mediates the optical feedback, but any fabrication method has finite accuracy making disorder ubiquitous, which induces light loss and results in hampered functionalities. Random lasers may overcome this limitation since they originate from the random multiple scattering of light induced by disorder~\cite{Cao1999,Genack2005,VanderMolen2007,Tureci2008,Gottardo2008,Wiersma2008,Kalt2009}. However, their multidirectional lasing emission, the lack of a broadband emission tunability, and the poor lasing mode confinement found in standard - diffusive - random lasers give rise to strong mode competition~\cite{Wiersma2008} and chaotic laser emission~\cite{Cahotic_effects}, and has so far limited their practical usability. Here we report on broadband-tunable random lasing in photonic-crystal waveguides in the regime of Anderson localization with low thresholds and small mode volumes. The strong confinement achieved by Anderson localization reduces the spatial overlap between lasing modes, thus preventing mode competition, which improves the laser output stability. Furthermore, a rich physical phase space is unraveled where the system can be driven in and out of the Anderson-localization regime by introducing gain or loss, thus demonstrating all-optical control of the lasing process. A statistical analysis of the data reveals the pathway towards minimizing the lasing threshold and mode volume by reducing the localization length, which provides unique possibilities of controlling and optimizing random lasing at the nanoscale.

The intrinsic disorder present in state-of-the-art photonic-crystal waveguides turns out to be sufficient to induce Anderson-localized cavities in the slow-light regime of waveguide propagation. Figure~\ref{random_lasing}a illustrates such spontaneous lasing cavities that are spatially distributed along the waveguide and spectrally bound to the slow-light regime \cite{Joannopoulos}. In the present experiment, quantum wells are embedded as gain medium in the photonic crystal membrane and are optically pumped. For low pump power, the strong absorption in the quantum wells damps interference and hence also Anderson localization, as illustrated in Fig.~\ref{random_lasing}b, while for increasing  excitation power, light amplification compensates the losses and stimulates the transition to the random lasing regime, as illustrated in Fig.~\ref{random_lasing}c. Figure~\ref{random_lasing}d displays the emission spectrum of a single Anderson-localized lasing mode below and above threshold, while Fig.~\ref{random_lasing}e shows the peak output emission intensity vs.~input excitation power. A distinct lasing threshold is observed, accompanied by a decrease of the cavity linewidth with excitation power that are the signatures of laser oscillation. Two important figures-of-merit of a nanolaser are the $\beta$-factor and the cavity mode volume, which are the fraction of spontaneous emission that couples to the lasing mode and the spatial extension of the lasing mode, respectively. They can be determined by modeling the laser input-output curves with micro-cavity semiconductor-laser rate equations~\cite{Gregersen_APL} (see Fig.~\ref{random_lasing}e). Details on the model can be found in the supplementary information (S.I.). For this particular lasing mode we obtain a $\beta$-factor of $0.31$, which interestingly is significantly larger than previous values reported in photonic-crystal lasers with quantum-well gain media~\cite{Altug2005}. A mode volume of $4.53\,(\lambda$/n)$^3$ is obtained, where $\lambda=1587\,\text{nm}$ is the lasing wavelength and $n=3.4$ is the refractive index of InGaAsP, which is consistent with the mode extent found for Anderson-localized modes in cavity quantum electrodynamics experiments~\cite{Smolka2011,Thyrrestrup2012}.

Determined by the size of the pumped area on the waveguide, multi-frequency lasing can be observed, as presented in Fig.~\ref{random_lasing}f and \ref{random_lasing}g. As shown in Fig.~\ref{random_lasing}f, the onset of multi-mode lasing tends to be sequential, i.e., with increasing excitation power the laser peaks grow one by one, each displaying the characteristic input-output lasing curve (see Fig.~\ref{random_lasing}g). No signature of mode competition is observed, which is attributed to the fact that the Anderson-localized lasing modes tend to be so strongly confined that different modes do not deplete the same gain region. In diffusive random lasers, on the contrary, the extended lasing modes imply that different modes overlap spatially, which induces strong mode competition and chaotic lasing emission~\cite{Cahotic_effects}. The improved stability of the multi-mode Anderson-localized random laser compared to diffusive random lasers may have significant advantages for mode-locked operation~\cite{Mode_locked}.

In diffusive random lasers, the lasing wavelength is primarily determined by the spectral peak of the gain medium~\cite{Cao1999} while only weak dispersion has been incorporated into the multiple-scattering medium enabling modest tunability of the random laser~\cite{Gottardo2008,Ramy2011}. A photonic-crystal waveguide, on the contrary, is highly dispersive, which implies that the Anderson-localized random lasers can be spectrally tuned over an unprecedented wavelength range by varying the photonic-crystal lattice, with a tunability that is only limited by the bandwidth of the quantum well gain medium (almost $300\,\text{nm}$). Figure~\ref{Tunability}a shows the spectral tuning of the lasing modes over $50\,\text{nm}$ when varying the ratio $r/a$, where $r$ is the hole radius and $a$ the lattice constant of the photonic-crystal waveguide. This broad tunability is due to a particular characteristic of Anderson-localized modes in photonic-crystal waveguides: the modes appear solely in the slow-light regime of the waveguide near the band-edge cutoff, where the density of optical states (DOS) associated with the waveguide mode is large~\cite{Garcia2010}.  Figure~\ref{Tunability}b displays how the high-DOS region of the waveguide can be tuned by varying the photonic-crystal parameters, establishing the tuning mechanism for the Anderson-localized modes.

Anderson localization is a random process and is described only by a statistical distribution of lasing parameters. The distribution of $Q$-factors at lasing threshold for all the modes observed in the experiment is shown in Fig.~\ref{Statistics}a, from which we can extract the localization length and the loss length~\cite{Smolka2011}. The localization length is the average distance between each random scattering event and we extract $\xi=6.0\,\micro\text{m}$ together with a loss length of $l=800\,\micro\text{m}.$ The latter is determined by out-of-plane scattering and absorption in the unpumped region of the quantum well, and is likely to be biased towards low loss since preferentially the very localized modes will lase. Since the localization length is significantly shorter than the sample length of $100\,\micro\text{m}$, the sample is found to be deeply in the Anderson-localization regime at lasing threshold. Figure~\ref{Statistics}b displays the distribution of lasing thresholds observed experimentally. An asymmetric threshold distribution with the long tail extending to high threshold values is obtained as predicted by theory~\cite{Apalkov2005}, in contrast to the symmetric threshold distributions observed for diffusive random lasers~\cite{Tulek2010}. The average threshold observed of $418\,\micro\text{W}$ is comparable to the typical threshold of standard photonic-crystal lasers pumped in a similar geometry\cite{Locar2002,Altug2005}, illustrating that Anderson-localized cavities may compete favorably with engineered nanocavity lasers that are hampered by imperfections. The $\beta$-factor and mode-volume distributions are plotted in Fig.~\ref{Statistics}c and Fig.~\ref{Statistics}d. Large $\beta$-factors ranging from $0.28$ to $0.38$ are found for all lasing modes in conjunction with small mode volumes ranging from 4 to 6 ($\lambda$/n)$^3$. The mode-volume distribution enables another estimate of the mode-length extension along the waveguide (see S.I.), which is plotted as the top horizontal axis in Fig.~\ref{Statistics}d. The average mode length of $4.3\,\micro\text{m}$ is very similar to the localization length ($\xi=6.0\,\micro\text{m}$) that is extracted from modelling the $Q$-factor distributions. This remarkable agreement confirms the validity of the two independent models applied for analyzing the experimental data, and illustrates that important microscopic parameters of Anderson-localized random lasers can be reliably extracted.

\begin{figure}
\centering
\includegraphics[width=\columnwidth]{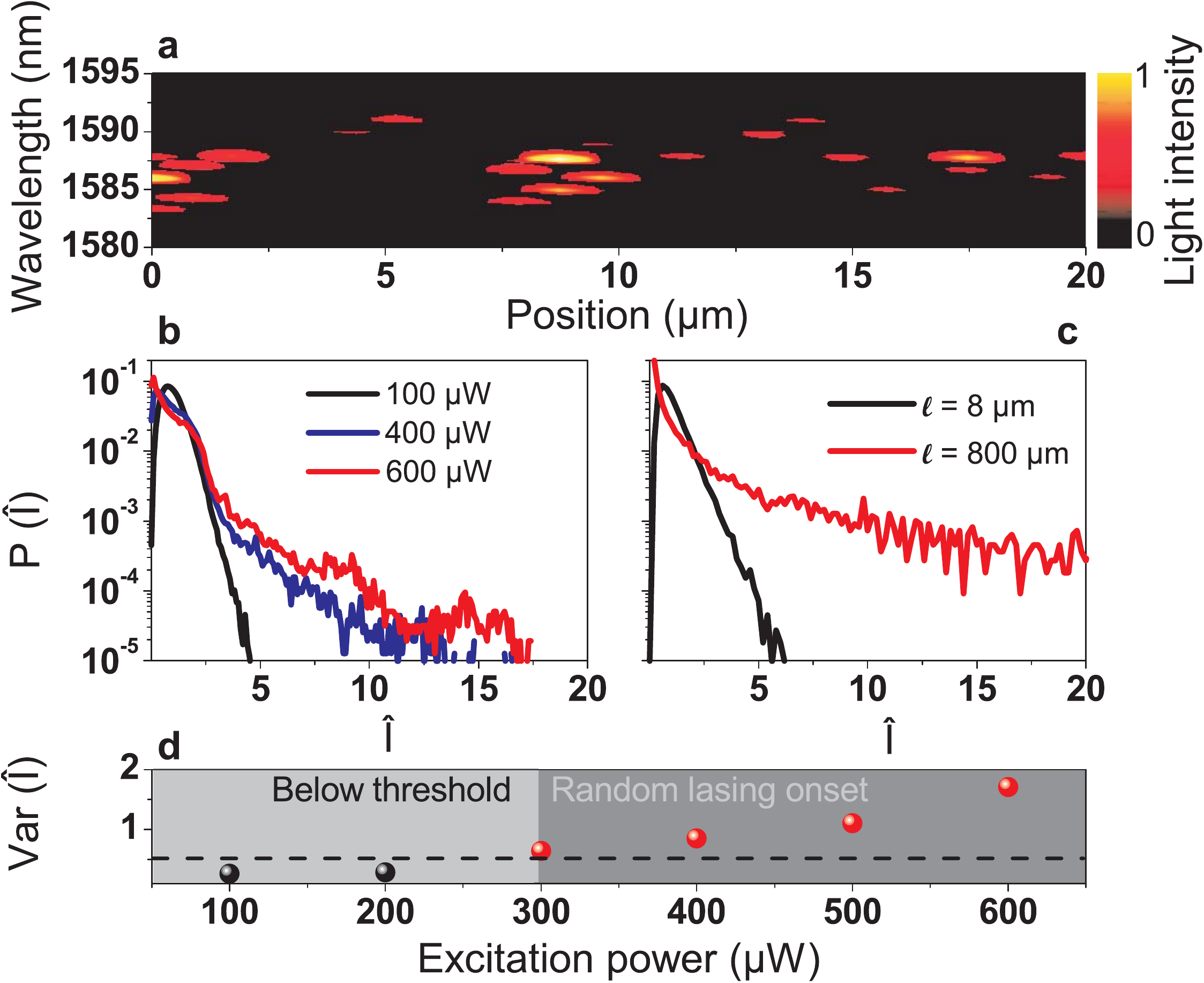}
\caption{\textbf{All-optical control of the cross-over to Anderson localization.} \textbf{a}, Photoluminescence intensity for an excitation power of \unit{500}{\micro\watt} collected while scanning the excitation and collection objective along a photonic-crystal waveguide. \textbf{b}, Intensity probability distribution for different excitation powers. \textbf{c}, Calculated intensity probability distribution with the model of Ref.~\citenst{Smolka2011} with a localization length of $\xi=6\,\micro\text{m}$ and two different loss lengths. $l=800\,\micro\text{m}$ corresponds to the loss length extracted from the $Q$-factor distributions measured at threshold. \textbf{d}, Variance of the normalized intensity as a function of the excitation power. The horizontal dashed line shows the localization criterion of $\mathrm{Var}(\widehat{I})=0.5$, and the vertical line indicates the excitation power at which the cavities start lasing.}
\label{Crossover}
\end{figure}

An important question is how to improve the lasing figures-of-merit in the case of random lasing. The ultimate nanolaser will ideally be thresholdless due to the $\beta$-factor approaching unity. The statistical analysis presented here offers insight to the governing parameters. Figures~\ref{Statistics}e,f and~\ref{Statistics}g,h show the measured lasing threshold and $\beta$-factor versus $Q$-factor and mode volume. The correlation between these parameters is quantified by the Spearman's rank correlation coefficient $\text{r}_{\mathrm{s}}$ (details in S.I.) that demonstrates how a high $Q$-factor and a small mode volume improve lasing performance. In Anderson-localized random lasers these cavity characteristics are directly linked to the localization length scaled to the sample length, which is a universal parameter determining all properties of light transport in the random medium. The key to improving the lasing performance of the Anderson-localized lasers even further would be to shorten the localization length, which may be feasible, e.g., by combining random disorder with long-range correlated order. Indeed, a shorter localization length would be beneficial in two ways since it would simultaneously increase the laser $Q$-factor and decrease the mode volume~\cite{Thyrrestrup2012}. However, unlike transport properties of random media, the Anderson-localized random lasers are not fully determined by universal parameters, and the non-universality of light emission in a random medium has recently been shown experimentally \cite{C0}. The current exploration of light emission and propagation in random media reveals the exciting prospects of disordered nanostructures for lasers and cavity quantum electrodynamics experiments.

Any amount of absorption or light leakage from the structure reduces the confinement due to Anderson localization. While inclusion of large amounts of gain inevitably modifies the properties of the underlying ordered structure~\cite{Jure}, moderate optical gain can compensate the effect of loss, thereby recovering the confinement of light. The effect of gain on Anderson localization can be investigated by extracting the probability distribution of the intensity fluctuations $P(\widehat{I} \equiv I/\langle I \rangle)$, where $\langle I \rangle$ denotes the intensity obtained after ensemble averaging over configurations of disorder. The intensity, $I$, is measured while raster scanning the sample, as shown in Fig.~\ref{Crossover}a (see S.I. for details), and displayed for different excitation powers in Fig.~\ref{Crossover}b. For low excitation power, the strong absorption from the quantum wells damps long interference paths resulting in the normal distribution of $P(\widehat{I})$ shown in Fig.~\ref{Crossover}b. When increasing the excitation power, light amplification compensates absorption, leading to Anderson localization and random lasing, for which $P(\widehat{I})$ displays the characteristic log-normal distribution of Anderson localization. Although no complete theory of Anderson localization in active media is yet available, the cross-over to localization can be qualitatively reproduced with a 1D multiple scattering model~\cite{Smolka2011} where the effect of gain is modeled by increasing the loss length, see Fig.~\ref{Crossover}c. The experimentally recorded variance of the intensity fluctuations, $\mathrm{Var}(\widehat{I}),$ is found to increase monotonically with excitation power, as shown in Fig.~\ref{Crossover}d. At the onset of random lasing, $\mathrm{Var}(\widehat{I})=0.85$ is recorded, which is larger than the threshold for Anderson localization in a 1D structure~\cite{Chabanov2000,Smolka2010} of $\mathrm{Var}(\widehat{I})=0.5$. The ability to manipulate loss and gain in a non-conservative 1D random system opens for the possibility of exploring the very rich parameter phase space that has been theoretically proposed~\cite{Yamilov2010}. The all-optical control of the cross-over from a lossy regime to Anderson localization offers exciting possibilities of fine-tuning the Anderson-localized modes.

Embedding light emitters in Anderson-localized random media provide a promising new pathway to enhanced light-matter interaction relying on the natural occurrence of cavities in random media rather than nano-engineering. The rich and intricate physics is currently being explored, but the prospective applications for low-threshold nanolasers and highly-efficient single-photon sources appear very attractive since the relevant figures-of-merit for these applications match engineered systems, with the added value that random structures could be much easier to fabricate. The fundamental limits of the disordered media are still to be explored, but the performance is likely to be improved even further by deliberately mixing the amount of order and disorder in the structures. An important issue for applications of an Anderson-localized nanolaser will be the ability to couple light from the randomly positioned laser to a well-defined single mode. This could easily be implemented on-chip by positioning another photonic-crystal waveguide, tailored to the non-localizing "fast light" regime, next to the Anderson-localizing waveguide, which would enable a highly efficient and stable multi-color laser with directional output.

\section{Acknowledgements}

We thank E.~Semenova for epitaxial growth. We furthermore gratefully acknowledge the Danish Council for Independent Research (Natural Sciences and Technology and Production Sciences), the European Research Council (ERC consolidator grant), and the Villum Foundation (NATEC center of excellence) for financial support.

\newpage

\section{supplementary information}

\subsection{Sample fabrication}

The semiconductor heterostructure used for the fabrication of the photonic-crystal waveguides was grown on a (100)-oriented semi-insulating indium phosphide (InP) substrate by metalorganic chemical vapor deposition (MOCVD). The structure consists of a 340-nm-thick $\text{In}_{0.77}\text{GaAs}_{0.503}\text{P}$ membrane, which is grown on a 100-nm-thick InP, 100-nm-thick InAlAs and 800-nm-thick InP sacrificial layer, incorporating 10 layers of $\text{In}_{0.75}\text{GaAs}_{0.86}\text{P}$ quantum well at the center with a thickness of 5 nm and a separation of 5 nm between adjacent quantum wells. The total refractive index of the structure is $n=3.4$. The photonic-crystal waveguides are fabricated by electron-beam lithography, reactive-ion etching, and wet etching using hydrochloric and hydrofluoride acid solutions. Samples with lattice constant $\textit{a}=380\,\text{nm}$, and a range of different hole radii $0.237\textit{a}<\textit{r}<0.263\textit{a}$ are fabricated. Finally, 340-nm-thick suspended membranes are formed by removing the sacrificial layer.

\subsection{Experimental setup and optical characterization}

Microphotoluminescence measurements are carried out at room temperature. A 120 fs pulsed Ti:sapphire laser operated at 800 nm with a repetition rate of 80 MHz is used for optical pumping. An excitation beam is focused to a spot with a diameter of $1.5\,\micro\text{m}$ on the surface of the sample using a $50\times$ microscope objective lens (numerical aperture: 0.65), and aligned to the photonic-crystal waveguides using piezoelectric nanopositioners. The photoluminescence signal is collected by the same objective lens within a diffraction-limited region in a wide wavelength range $1500\,\text{nm} < \lambda < 1600\,\text{nm}$, dispersed by a 300 mm grating spectrograph with a spectral resolution of 0.1 nm, and detected using a liquid-nitrogen-cooled InGaAs CCD (charge-coupled device) camera.

We measure a total of 25 input-output curves by varying the excitation and collection position along the photonic-crystal waveguide. The laser threshold is obtained by extrapolating each input-output curve in a linear scale to zero power. The $Q$-factor is extracted by fitting the cavity mode with a Lorentzian at the threshold power. Two out of the 25 input-output curves show the cavity mode also below the laser threshold (Fig.~\ref{random_lasing}b shows one example). For the rest of the measurements, the excitation power at which the cavity mode appears matches the threshold power extracted from the input-output curve fitting. Based on this fact, we can increase the statistics of threshold and $Q$-factor distributions by gradually increasing the excitation power and recording the power at which the mode appears and its corresponding spectrum. By doing so, 25 extra data points for the laser threshold and $Q$-factor are added without recording the full input-output curve. The full data set is plotted in Fig.~\ref{Statistics}c and ~\ref{Statistics}d.

The intensity probability distribution $P(\widehat{I})$ is measured by collecting the intensity $I^{\lambda}_{x,y}$ while scanning the excitation and collection objective along the photonic-crystal waveguide at each spatial position $(x,y)$ with a spatial binning size of $0.25\,\micro\text{m}$ and at different wavelengths with a binning size of $1 \,\text{nm}$. Subsequently, an average over the wavelength range $1580\,\text{nm} - 1595\,\text{nm}$ is performed.

\subsection{Rate-equation analysis}

We use a modified semiconductor-laser rate-equation model for the carrier density $N$ and the photon density $P$ of the random laser~\cite{C1995_S}
\begin{align}
&\frac{\mathrm{d}N}{\mathrm{d}t} = R_\mathrm{p} - BN^{2}-\frac{\Gamma^\mathrm{G}}{\Gamma}G(N)P-AN-CN^{3},    \\
&\frac{\mathrm{d}P}{\mathrm{d}t} = \Gamma\beta BN^{2}+\Gamma^\mathrm{G}G(N)P-\frac{P}{\tau_\mathrm{P}},
\end{align}
where $R_\mathrm{p}$ is the pulsed injected carrier density with the temporal width of the excitation laser pulse, $B$ is the bimolecular recombination coefficient of the unstructured material that we assume here to be unchanged by the photonic crystal structure, and $A$ and $C$ are the non-radiative coefficients corresponding to surface recombination and Auger processes, respectively. $\Gamma$ and $\Gamma^\mathrm{G}$ are the conventional and generalized confinement factors, $G(N)$ = c/$n_\mathrm{eff}$$G_{0}$($N$-$N_\mathrm{tr}$) is the gain coefficient, and $\tau_\mathrm{P}$ is the photon lifetime obtained from the measured $Q$-factor at the laser threshold (the remaining parameters are defined in Table \ref{table:mechanical characteristics}).\ The formalism takes into account Purcell enhancement of the spontaneous and stimulated emission~\cite{Gregersen_APL_S} including the fact that the quantum well electronic density of states  is very broadband compared to the cavity bandwidth~\cite{Suhr2010_S}. Furthermore, in the present experiment where only a small volume is optically pumped, the cavity mode volume is comparable to the volume of the gain material. The conventional optical confinement factor, $\Gamma = V_\mathrm{a}/V$, where $V_\mathrm{a}$ and $V$ are the active material and modes volume, respectively, does not take into account the spatial-modal overlap profile with the gain, and a generalized confinement factor~\cite{Mock 2010_S} should be used. The mode volume $V$ and the generalized confinement factor $\Gamma^\mathrm{G}$ are defined as
\begin{align}
&V=\frac{\int{\epsilon_\mathrm{r}(\textbf{r})|\textbf{E}_\mathrm{c}(\textbf{r})|^{2}\mathrm{d}^3\mathbf{r}}}{\epsilon_\mathrm{r}(\textbf{r}_{0})|\textbf{E}_\mathrm{c}(\textbf{r}_{0})|^{2}} \\
&\Gamma^{G} = \frac{\int_{V_\mathrm{a}}\epsilon_\mathrm{r}(\textbf{r})|\textbf{E}_\mathrm{c}(\textbf{r})|^{2}\mathrm{d}^3\mathbf{r}}{\int\epsilon_\mathrm{r}(\textbf{r})|\textbf{E}_\mathrm{c}(\textbf{r})|^{2}\mathrm{d}^3\mathbf{r}}
\end{align}
where $\textbf{r}_{0}$ is the position of the anti-node of the cavity field, $V$ is the mode volume of the Anderson-localized laser mode, $V_{a}$ is the active volume covered by the carriers after excitation and subsequent diffusion, and $\textbf{E}_\mathrm{c}(\textbf{r})$ is the cavity-mode electric profile. In order to reduce the number of free  parameters, we simplify the expression of $\Gamma^\mathrm{G}$ by assuming that the cavity field is uniformly distributed spatially with the amplitude $|\textbf{E}_\mathrm{c}(\textbf{r})|^{2}$=$\frac{1}{2}$$|\textbf{E}_\mathrm{c}(\textbf{r}_{0})|^{2}$ and $\epsilon_\mathrm{r}(\textbf{r})=\epsilon_\mathrm{r}(\textbf{r}_{0})$, leading to
\begin{equation}\
\Gamma^\mathrm{G}=\frac{V_\mathrm{a}^\mathrm{G}}{2V}
\end{equation}
where $V_\mathrm{a}^\mathrm{G}$ is the overlap volume between the cavity mode and the active volume.
\begin{table}[ht]
\caption{Material parameters for the semiconductor-laser rate-equation~\cite{C1995_S}}
\centering
\begin{tabular}{c}
\hline\hline
\hline
Surface recombination rate ($A$) = $2.5\times 10^{9}$ $\mathrm{s^{-1}}$ \\
Bimolecular recombination rate ($B$) = $10^{-16}$ $\mathrm{m^{3}/s}$    \\
Auger non-radiative recombination rate ($C$) = $8\times 10^{-41}$ $\mathrm{m^{6}/s}$ \\
Transparency carrier density ($N_\mathrm{tr}$) = $10^{24}$ $\mathrm{m^{-3}}$ \\
Gain coefficient ($G_{0}$) =  $2.5\times 10^{5}$ $\mathrm{m^{2}}$  \\
Effective refractive index ($n_\mathrm{eff}$)= 2.8  \\
\hline\hline
\end{tabular}
\label{table:mechanical characteristics}
\end{table}

\begin{figure}[b!]
\centering
\includegraphics[width=\columnwidth]{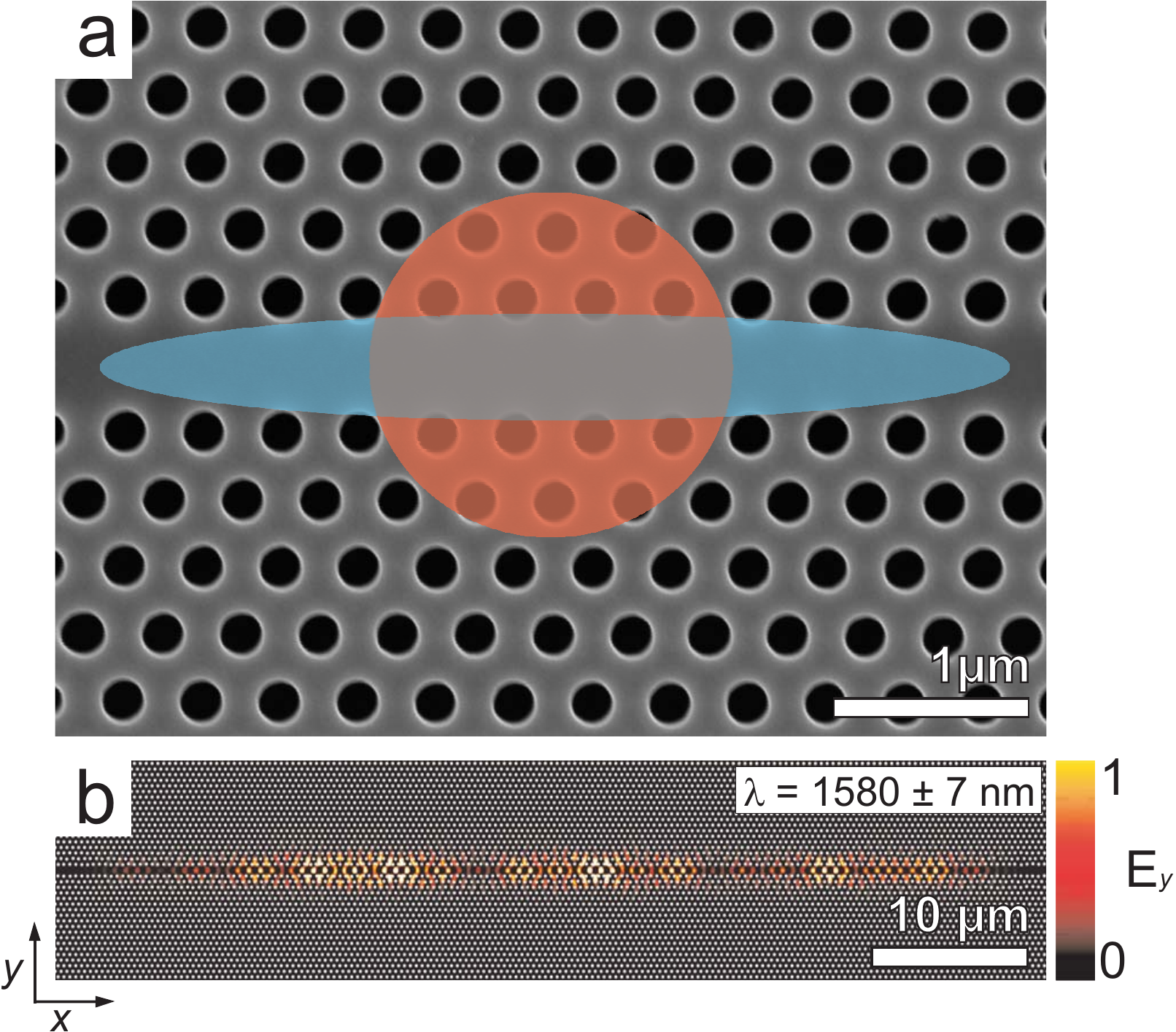}
\caption{\textbf{Schematic illustration of the active area and cavity mode area.} \textbf{a}, Scanning electron micrograph of a photonic-crystal waveguide. The red area indicates the optically excited and thus active area, $V_{a}$, and the blue area represents the area of a random laser mode, determining $V$. The overlap between them gives the effective active overlap, $V_\mathrm{a}^\mathrm{G}$. \textbf{b}, 2D finite-difference time-domain calculations of the $y$-component of the electromagnetic field in a disordered photonic-crystal waveguide with $\textit{r}/\textit{a} = 0.28$ and $\delta=3\%$. Strongly confined Anderson-localized modes are found in the slow-light regime after exciting the structure with a broadband source centered at 0.26$\textit{a}/\lambda$ with a bandwidth of 0.02$\textit{a}/\lambda$. The intensity of the $y$-polarized electric field is plotted on top of the simulated structure. }
\label{supplementary}
\end{figure}

We estimate this overlap volume as follows. The size of the active area is determined by the excitation beam, with a $1.5\,\micro\text{m}$ diameter (red area in Fig.~\ref{supplementary}a). The carrier diffusion distance is $l=\sqrt{D\tau}=0.9\,\micro\text{m}$, by assuming $D=8.0\,\text{cm}^{2}\text{/s}$ as the carrier diffusion coefficient and $\tau=1\,\text{ns}$ as the carrier lifetime. We neglect the carrier diffusion distance in the estimation of the active area since the carrier lifetime is expected to be much shorter above threshold giving rise to a much shorter diffusion distance. We assume the Anderson-localized modes to be confined transversally by the effect of the 2D photonic-crystal gap (blue area in Fig.~\ref{supplementary}a). Therefore, a total overlap length of $1.5\,\micro\text{m}$ is determined by the active area, while the overlap width of $2a \sin \frac{\pi}{3}-2r=458\,\text{nm}$ is estimated by the width of the photonic-crystal waveguide. The overlap height is given by the total thickness of 10 QW layers ($\mathrm{5nm}\times10$). In summary, we estimate
\begin{align}
&V_\mathrm{a}^\mathrm{G}=(1500\times458\times50)~\mathrm{nm^{3}} \\
&V_\mathrm{a}=((\pi\times750^{2}-14\times\pi\times100^{2})\times50)~\mathrm{nm^{3}}
\end{align}
where the area of 14 holes is subtracted from the excitation spot area to account for the final active area (see Fig.~\ref{supplementary}a).

Finally, we estimate the cavity mode length by dividing the mode volume obtained from the semiconductor-laser rate-equation model with an effective transverse mode area $\mathrm{A_{eff}}$. To calculate $\mathrm{A_{eff}}$, we assume the Anderson-localized modes to be confined transversally. The width of the cavity mode can be calculated from finite-difference time-domain simulations of a photonic-crystal waveguide where disorder is induced by randomly varying the hole positions according to a Gaussian distribution with a standard deviation $\delta=3\%$ relative to the lattice spacing (see Fig.~\ref{supplementary}b) and the thickness of the cavity mode is assumed to be the same as the membrane. By averaging over one unit cell along the waveguide we obtain $\mathrm{A_{eff}} = 0.11\,\micro\text{m}^{2}$.

\subsection{Investigating correlations of random laser parameters}

In order to investigate how to optimize the performance of a random laser, we estimate the Spearman's rank correlation coefficient, $\text{r}_{\mathrm{s}}$, between the different laser parameters. In particular, we focus on the dependence of the laser threshold $P_{\mathrm{th}}$ and the $\beta$-factor on the Anderson-localized mode properties, i.e., the $Q$-factor and mode volume. $\text{r}_{\mathrm{s}}$ is a non-parametric measure of the statistical correlation between two variables, regardless the particular functional dependence between them \cite{Spearman}. The Spearman correlation varies from 1 (fully correlated) to -1 (fully anti-correlated).

\end{document}